\newif\ifboo \boofalse

\def\Review#1{\boofalse{\it #1},}
\def\Name#1{{\sc #1},}
\def\Vol#1{\ifboo Vol. {\bf #1}\else{\bf #1}\fi}
\def\Year#1{\ifboo #1\else(#1)\fi}
\def\Book#1{\bootrue{\it #1},}
\def\Page#1{\ifboo {\rm p. #1}\else{\rm #1}\fi}

\documentstyle[prb,aps]{revtex}
\input epsf
\begin{document}

\title{Non-Gaussian distribution of nearest-neighbour Coulomb 
 peak spacings in metallic single electron transistors}

\author{M. Furlan$^{1,2}$, T. Heinzel$^{1}$, 
     B. Jeanneret$^{2}$, S. V. Lotkhov$^{3}$
     and K. Ensslin$^{1}$}

\address{
     $^{1}$ Solid State Physics Laboratory, ETH Z\"urich,
     CH-8093 Z\"urich, Switzerland \\
     $^{2}$ Swiss Federal Office of Metrology OFMET,
     CH-3003 Bern-Wabern, Switzerland\\
     $^{3}$ Physikalisch-Technische Bundesanstalt PTB, 
     D-38116 Braunschweig, Germany}


\maketitle

\begin{abstract}
The distribution of nearest-neighbour spacings of Coulomb blockade 
oscillation peaks in normal conducting aluminum single electron 
transistors is found to be non-Gaussian.  A pronounced tail to reduced 
spacings is observed, which we attribute to impurity-specific 
parametric charge rearrangements close to the transistor.  Our 
observation may explain the absence of a Wigner-Dyson distribution in 
the experimental nearest-neighbour spacing distributions in 
semiconductor quantum dots.
\end{abstract}

\section{Introduction}
A single electron transistor (SET) consists of a small conductive 
island, coupled to two leads via tunnel barriers, and a nearby gate 
used to tune the electrochemical potential of the 
island.\cite{SETrev.r} The Coulomb blockade, characterized by the 
charging energy $E_{\mathrm{C}}$ needed to add a single electron to 
the island, governs the electronic properties of such devices.  This 
leads to the observation of pronounced conductance oscillations, 
commonly denoted as Coulomb blockade (CB) oscillations, as a function 
of the gate voltage $V_{\mathrm{g}}$.  These effects are of 
electrostatic origin and can be analyzed in a purely classical 
picture.  However, a variety of additional effects can be studied in 
SETs, depending on the material they are made of.  In superconducting 
islands, for example, Cooper pair formation leads to significant 
modifications of the device characteristics.  SETs can also be 
realized in two-dimensional electron gases residing in semiconductor 
hosts such as Si MOSFETs or Ga(Al)As heterostructures.\cite{qdot.r} 
Discrete energy levels and phase coherence effects superimposed on the 
Coulomb blockade can be observed.  Such devices, also known as 
`quantum dots', have therefore become model systems to investigate 
numerous distinct effects.  Broad attention has recently been paid to 
experiments measuring the distribution of nearest-neighbour spacings 
(NNS) of the CB oscillation peaks in quantum dots.  From random-matrix 
theory calculations, the NNS distribution is expected to obey 
Wigner-Dyson statistics.\cite{RMT.r} However, the experimentally 
observed distributions differ significantly from the random-matrix 
theory predictions.\cite{expNNS.r,simmel2.r,maurer.r} In order to 
separate classical charging effects from quantum mechanics, it is 
generally accepted to use a constant-interaction (CI) 
model,\cite{qdot.r} which assumes that the electrostatics of the 
system is invariant under a change of the charge on the island by 
integer multiples of the elementary charge $e$.

In this paper, we report on NNSs of CB peaks in metallic, 
\textit{i.e.} purely electrostatic or `classical' SETs.  In a simple 
picture appropriate for metallic devices, one would expect to observe 
constant peak spacings $\Delta V_{\mathrm{g}} = e/C_{\mathrm{g}}$ 
(with a distribution broadened by thermal fluctuations only), where 
$C_{\mathrm{g}}$ is the capacitance between island and gate.  However, 
we observe a strongly asymmetric NNS distribution with a pronounced 
tail to small peak separations.

\section{Experiments and results}
\begin{figure}
   \bigskip
   \centerline{\epsfxsize=17cm \epsfbox{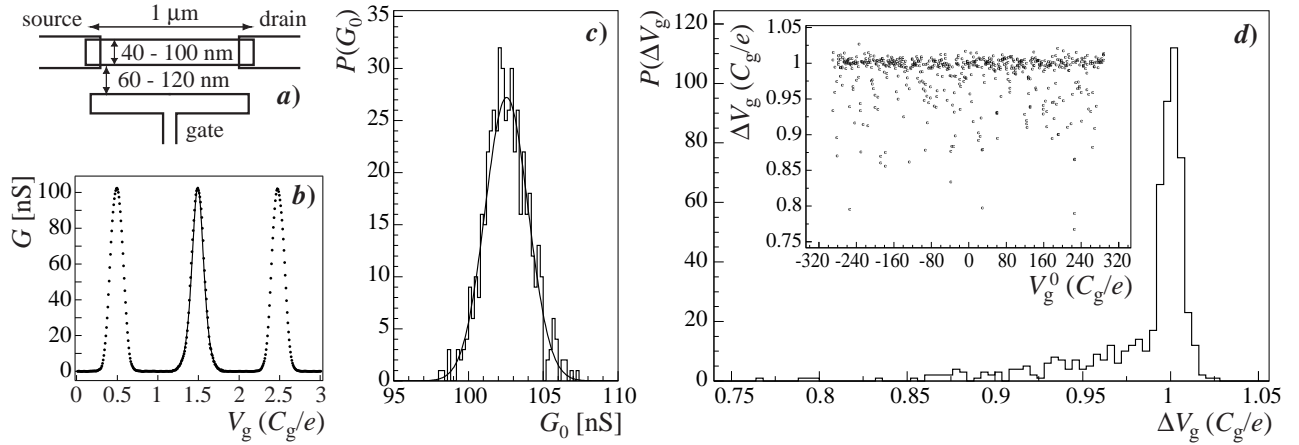}}
   \medskip
   \caption{\textit{a}) Schematic layout of the investigated
   SET devices. The tunnel barriers are formed between the 
   overlap of island and electrodes. 
   \textit{b}) Typical measured conductance oscillations (dots). 
   Perfect corresponance between experimental data and theoretical 
   fit (see main text) is shown for the middle peak (solid line).
   The distributions of CB peak amplitudes (together with a Gaussian 
   fit) and NNSs are shown in \textit{c}) and \textit{d}), 
   respectively. The inset in \textit{d}) shows 
   $\Delta V_{\mathrm{g}}$ as a function of the peak position.}
   \bigskip
  \label{fig1.f}
\end{figure}

We have measured high quality Al/AlO$_{\rm{x}}$/Al SETs written by 
electron beam lithography and fabricated by standard two angle 
evaporation technique,\cite{Fulton.r} with intermediate room 
temperature oxidation of the first layer (oxygen pressure of 2.5 mbar 
for 20~min) to develop the tunnel barriers.  The substrate was silicon 
covered by 600~nm thermally grown SiO$_{\rm{x}}$.  The design of the 
SETs is drawn schematically in fig.~\ref{fig1.f}~\textit{a}).  The 
typical parameters of the devices were $R_{\mathrm{t}} = 1 \ldots 
10$~M$\Omega$ for the tunnel resistances, $C_{\mathrm{j}} = 40 \ldots 
200$~aF and $C_{\mathrm{g}} \approx 50$~aF for the junction and gate 
capacitances, respectively.  The measurements were performed in a 
dilution refrigerator at temperatures down to 5 mK, whereas the 
effective electron temperature of the devices was determined to be 
45~mK (as deduced from the thermal smearing of the charge occupation 
number in an electron box.\cite{ebox.r}) The device IV-characteristics 
are well understood on the basis of `orthodox theory' calculations, 
taking into account also non-equilibrium effects and the influence of 
the electromagnetic environment.\cite{lt22miha.r} Measurements were 
performed over periods of several days.  The devices were highly 
stable for constant voltages applied, showing no drifts or spontaneous 
jumps for days.  We attribute this stability to the very slow device 
cooling of about 1 day, allowing the impurities to be frozen in their 
lowest, most stable state.  The measured $1/f$ noise was identified as 
dominant SET input noise due to background charge fluctuation, being 
of magnitude comparable with typical noise figures ($\approx 10^{-4} 
e/\sqrt{\mathrm{Hz}}$ at 10 Hz) reported for other metallic 
SETs.\cite{SETnoise.r} A magnetic field of $B = 1 \ldots 4$~T was 
applied to suppress superconductivity.

The DC current through the devices was measured as a function of bias 
and gate voltages $V_{\mathrm{b}}$ and $V_{\mathrm{g}}$, respectively.  
Due to the DC measurement technique we took large sets of sufficiently 
dense points by variation of the voltages in the ranges 
$|V_{\mathrm{b}}| \leq \frac{1}{4} E_{\mathrm{C}}/e$ and 
$|V_{\mathrm{g}}| \leq 1$~V. The latter corresponds to a difference 
of some hundred electrons on the island.  The data was analyzed by 
fitting the conductance peaks with 
$G(V_{\mathrm{g}}) = \frac12 G_{\mathrm{0}} (\delta V_{\mathrm{g}} / w) 
/ {\mathrm{sinh}} (\delta V_{\mathrm{g}} / w)$,\cite{kulik.r} 
where $\delta V_{\mathrm{g}} = |V_{\mathrm{g}}^0 - V_{\mathrm{g}}|$, 
yielding amplitude $G_{\mathrm{0}}$, width 
$w = (k_{\mathrm{B}}T / 2 E_{\mathrm{C}}) (e / C_{\mathrm{g}})$ 
and position $V_{\mathrm{g}}^0$.  
A partial trace of typical CB oscillations is shown in 
fig.~\ref{fig1.f}~\textit{b}) 
together with a theoretical curve fitting.
As expected for our devices, the amplitudes are 
found to be constant over 
the entire $V_{\mathrm{g}}$ range, with a standard deviation of 
typically $1.5\%$, as shown in fig.~\ref{fig1.f}~\textit{c}).
However, the distribution $P(\Delta V_{\mathrm{g}})$ of NNS values 
$\Delta V_{\mathrm{g}}(n) = V_{\mathrm{g}}^0 (n+1) - V_{\mathrm{g}}^0 
(n)$, where $n$ is the peak index, is not Gaussian but shows a 
significant number of events with reduced values, \textit{cf.}
fig.~\ref{fig1.f}~\textit{d}).  
The NNS distributions were quantitatively independent of 
$V_{\mathrm{b}}$ variations.  The 8 samples investigated all showed 
similar behaviour.  The main peak in $P(\Delta V_{\mathrm{g}})$, 
containing the majority of the events ($\approx 60 \ldots 80~\%$ for 
different samples), fits well to a Gaussian, whose
width  scales linearly with temperature.

We should mention that samples cooled at a much faster rate typically 
show strongly enhanced noise levels.  Consequently, measurements of CB 
peak statistics with such devices yielded significantly broadened NNS 
distributions (not shown).

\begin{figure}
   \bigskip
   \centerline{\epsfxsize=12cm \epsfbox{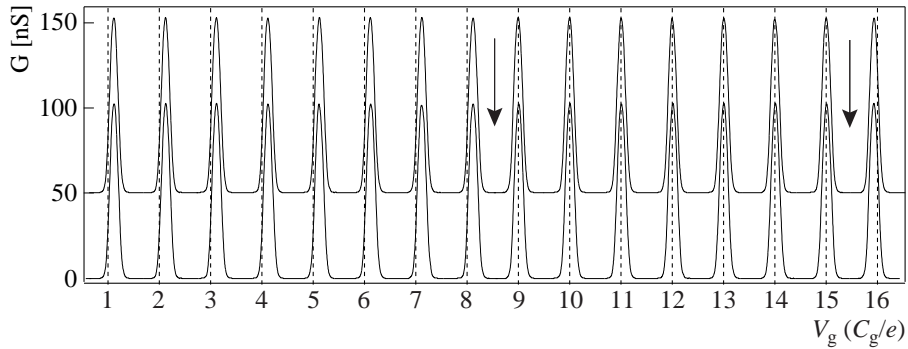}}
   \medskip
   \caption{Two traces of measured CB oscillations taken from 
            consecutive $V_{\mathrm{g}}$ scans swept in one   
            direction (the second trace is offset by 50 nS). 
            Two reduced NNSs in each trace are observed (denoted by 
            arrows), occurring at the same $V_{\mathrm{g}}$ positions 
            for both scans. The vertical dashed lines correspond to 
            a perfectly constant CB peak spacing of 
            $\Delta V_{\mathrm{g}} = e / C_{\mathrm{g}}$.}  
   \bigskip
   \label{fig2.f}
\end{figure}

In order to investigate reproducibility of the reduced NNS events, we 
have performed measurements on a smaller $V_{\mathrm{g}}$ range, where 
only very few NNSs with reduced values are detected.  
Figure~\ref{fig2.f} shows traces with 16 conductance peaks each, taken 
from two consecutive $V_{\mathrm{g}}$ scans in the same direction.  
Both traces show two shifts in $\Delta V_{\mathrm{g}}$ at the same 
positions.  It has been found in general that the position range where 
NNSs significantly smaller than the mean value $\langle \Delta 
V_{\mathrm{g}} \rangle$ occur, is well reproduced as a function of 
$V_{\mathrm{g}}$.  In addition, a clear difference in low $\Delta 
V_{\mathrm{g}}$ positions between up and down scans was observed, 
suggesting a hysteretic behaviour.  More details on the 
reproducibility and the hysteresis effect will be published 
elsewhere.\cite{chalmers99.r}

The CB peak position fluctuations do not show significant correlation 
as a function of $V_{\mathrm{g}}$.  The standard deviation of the 
$n$-th neighbour peak spacings \cite{maurer.r} is very closely 
proportional to $\sqrt{n}$, which is expected for uncorrelated events.  
We could not find any specific periodicity from a Fourier analysis of 
the CB peak position spectrum either.

The measured noise was essentially proportional to the SET gain 
$\mathrm{d} I / \mathrm{d} V_{\mathrm{g}}$, \textit{i.e.} dominated by 
device input noise.  In very few cases a significantly increased noise 
level was observed, with a non-zero correlation with events of reduced 
CB peak width.  This is attributed to the well-known dynamic switching 
of background charges (`random telegraph noise', RTN) for certain 
$V_{\mathrm{g}}$ values close to the fluctuator 
threshold.\cite{RTNrev.r} Considering the rare occurrence of 
correlated excess noise with a non-average NNS, we conclude that the 
fluctuators producing dynamic noise are not primarily responsible for 
the observed reduction of NNSs.

In addition, we should emphasize that the measured fluctuation 
distributions did not depend on the absolute $V_{\mathrm{g}}$ range 
considered (\textit{cf.} also inset in fig.~\ref{fig1.f}~\textit{d}) ).

\section{Discussion}
\begin{figure}
   \bigskip
   \centerline{\epsfxsize=16cm \epsfbox{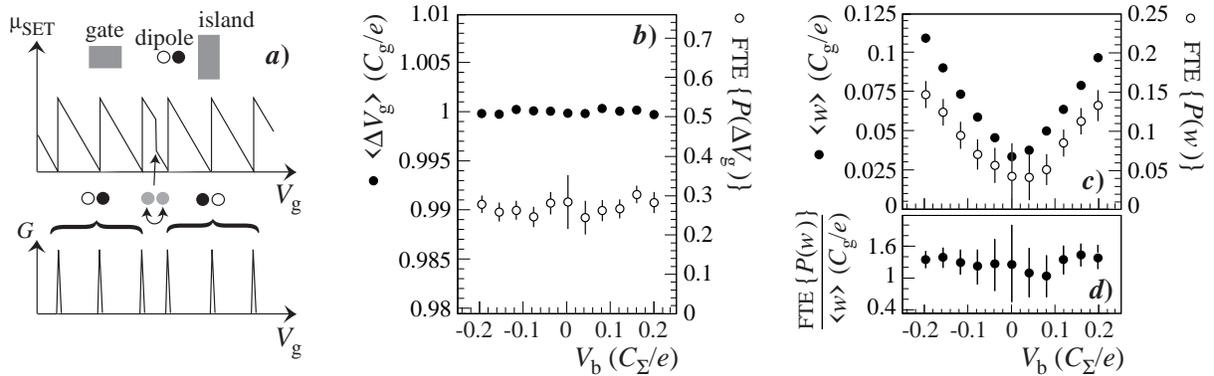}}
   \bigskip
   \caption{\textit{a}) Scheme of a dipole in between 
            island and gate, switching due to variation of 
            $V_{\mathrm{g}}$ with the consequence of a discontinuous 
            change of the island potential and a shift of the device 
            characteristics in $V_{\mathrm{g}}$ axis. 
            Plot \textit{b}) shows the main peak position 
            $\langle \Delta V_{\mathrm{g}} \rangle$ and the fraction 
            of the tail events (FTE) of the experimental 
            $P(\Delta V_{\mathrm{g}})$ distribution as a function of 
            $V_{\mathrm{b}}$ (full dots and open circles,
            respectively). The corresponding  measurements for the 
            distribution of the CB peak widths $P(w)$  
            are shown in \textit{c}).
            The ratio of the results in \textit{c}) is drawn 
            in \textit{d}).}
   \bigskip
   \label{fig3.f}
\end{figure}

Based on the theoretically expected behaviour of our SET transistors 
and the experimental results discussed above, we explain the 
observations with discontinuous switching of two-level tunnelling 
systems (TLTS),\cite{RTNrev.r} where the displacement of a single 
charge modifies the transistor island potential.  
Figure~\ref{fig3.f}~\textit{a}) 
shows schematically how such switching events can explain the 
systematic occurrence of reduced NNSs.  Consider a TLTS in a 
metastable state, located in between the gate and the SET island.  
Exceeding a particular $V_{\mathrm{g}}$ threshold, a charge 
rearrangement can be induced in the TLTS. In response to this, the 
electrochemical potential $\mu_{\mathrm{SET}}$ of the SET island, 
which is usually tuned 
continuously by $V_{\mathrm{g}}$, experiences a sudden jump in the 
same direction as the $V_{\mathrm{g}}$ variation, independent of the 
scan direction.  Consequently, a smaller $\Delta V_{\mathrm{g}}$ is 
needed in order to change the island occupation number by one.  Within 
this picture, the tail in the NNS distribution reflects the spatial 
and energetical distribution of TLTSs in some region between the 
island and the gate electrode.  The adjustment of a dipole following 
the variation of an electric field is equivalent to the picture of 
introducing a medium with increased dielectric constant, increasing 
$C_{\mathrm{g}}$ and decreasing $\Delta V_{\mathrm{g}}$.

Assuming a simple system of an electron switching locally between two 
sites, the measured CB peak spacing statistics $P(\Delta 
V_{\mathrm{g}})$ can be analyzed considering spacial distribution and 
type of such TLTSs.  According to electrostatic dipole 
calculations \cite{imagecharge.r} we derive the 
position of the electron and its displacement which allow a 
variation of island charge on the order of $10\%$ (as in our 
experiments): an electron located very close to the island ($\leq$~1~nm) 
and facing the gate electrode requires a displacement (radially away 
from the island) by $2\ldots4$~nm.  A process of a charge displacement 
by a few nanometres is very well consistent with other studies on 
charge trapping.\cite{RTNrev.r}  However, by slightly increasing the 
electron's distance from the island, the necessary displacement quickly 
grows to length scales for which the observed reproducibility of TLTS 
switching becomes very unlikely.  The largest electric fields are 
found between island and gate electrode, whereas the field is shielded 
or strongly reduced elsewhere, thereby reducing the trap switching 
effect to a negligible level.  This explains the asymmetry of 
$P(\Delta V_{\mathrm{g}})$ with a tendency to lower values.  Hence, 
the shape of the $P(\Delta V_{\mathrm{g}})$ distribution is determined 
by geometry and materials of the device and the surroundings.

SET transistors are known to be the best electrometers to date, with a 
sensitivity to charge variations by a small fraction of the electron 
charge $e$.  In contrast to the random dynamic background charge 
fluctuations, the TLTSs in our case are highly stable in time, 
depending (in first approximation) on electrical potential variations 
only.  The thermal activation energy is typically much larger than our 
temperature range investigated.  The reproducibility of the effects 
suggests a well-defined system of individual charge traps with 
negligible interaction.  Defects, acting as charge traps, may 
particularly reside in oxides, at semiconductor heterointerfaces, or 
generally at any disordered interface or lattice.  The low-frequency 
noise in metallic SET transistors is commonly attributed to background 
charge fluctuations mainly in the substrate, with a small probability 
of traps in the tunnel junctions.\cite{SETnoise.r}
A few studies on RTN have also been 
reported for semiconducting nanostructure devices.\cite{dotnoise.r}

Switching of background charges can be detected directly with the SET 
provided the device is in a sensitive state of non-zero gain, 
\textit{i.e.} within a conductance peak.  In our case, the switchings 
predominantly reduce the width of the peak.  Under low bias 
conditions, most of these events occur in between the CB peaks and are 
not seen in the peak width.  However, we can increase the detection 
range of the peaks for TLTS switching by making them wider, 
\textit{e.g.} by increasing the bias voltage.  The CB peak widths $w$ 
show a distribution $P(w)$ qualitatively similar to $P(\Delta 
V_{\mathrm{g}})$.  The peak width in $P(w)$ reflects the temperature 
of the system.  As plotted in fig.~\ref{fig3.f}~\textit{b}), the 
position of the main $P(\Delta V_{\mathrm{g}})$ peak shows no 
variation with $V_{\mathrm{b}}$, and the fraction of the tail events 
(FTE) in $P(\Delta V_{\mathrm{g}})$ also remains constant within 
experimental errors.  On the other hand, the $P(w)$ main peak position 
has a thermally broadened minimum at $V_{\mathrm{b}} = 0$ and 
increases almost linearly with increasing $|V_{\mathrm{b}}|$, 
\textit{cf.} fig.~\ref{fig3.f}~\textit{c}).  The FTE of the $P(w)$ 
distribution behaves proportionally to $\langle w \rangle$, which is 
indicated by their ratio in fig.~\ref{fig3.f}~\textit{d}).  This 
confirms that the discontinuous jumps in $V_{\mathrm{g}}$ are 
uniformly distributed along the $V_{\mathrm{g}}$ axis, independent of 
the state of the SET transistor, \textit{i.e.} whether it is in the CB 
regime or not.  Apparently, the positions of the jumps depend on the 
gate potential only.  Our arguments are further supported by 
correlated low tail events between the $P(\Delta V_{\mathrm{g}})$ and 
$P(w)$ distributions.

On one hand, our experimental results reveal important information for 
the understanding of charge fluctuation mechanisms in nanostructures, 
hopefully leading to an improvement of reliable and stable devices.  
This is particularly crucial for developments like quantum computing, 
ultra-low noise electrometers or metrological applications.  We have 
demonstrated the possiblity to detect directly discrete fluctuations 
of the background charge configuration, allowing a quantitative 
characterization of substrates or other dielectrics of interest.

On the other hand, we wish to emphasize in particular the impact of 
our results on the lively discussion on CB peak statistics in 
semiconductor quantum dots.  Our experiments show that even in the 
absence of single particle energy levels on the SET island, the NNS 
distribution can deviate significantly from a Gaussian, since it is an 
intrinsic feature of the island to react with high sensitivity to 
background charge rearrangements.  So far, all experiments studying 
the NNS distribution of quantum dots have been performed by measuring 
CB oscillations as a function of a gate 
electrode.\cite{expNNS.r,simmel2.r,maurer.r} The observed peak 
spacings in $V_{\mathrm{g}}$ have been corrected using the CI model.  
However, the remaining NNS distribution does contain the modification 
of level spacings as a consequence of rearrangements in the random 
background charge configuration and cannot solely be attributed to the 
energy spectrum of the quantum dot.  The charge and potential 
distribution in a two-dimensional electron gas (2DEG) is known to be 
fairly inhomogeneous and sensitive to even small perturbations of 
electromagnetic field.\cite{2DEGchaos.r} Furthermore, single electron 
charging effects among isolated regions due to non-uniform potential 
distribution in a 2DEG have recently been observed.\cite{cobden2.r} 
Consequently, charge sensitive nanodevices made of semiconducting 
structures may reveal a significantly modified behaviour due to 
charging effects, of the origin described above.  In terms of our 
model explanations, it can be easily understood, \textit{e.g.}, why 
Simmel \textit{et al.} \cite{simmel2.r} observe much broader NNS 
distributions in quantum dots defined in Si MOSFETs than those 
distributions observed in Ga(Al)As heterostructures, since there are 
more traps in SiO$_{2}$ than in heterostructures grown by molecular 
beam epitaxy.

In order to go beyond the CI model, we therefore suggest that the 
charge rearrangements in the vicinity of the quantum dot should be 
measured independently.  In detail, one could define a metallic 
`control' SET on top of a quantum dot, which is used to correct each 
individual peak spacing of the quantum dot for the charge fluctuations 
in the environment.

In summary, we have measured non-Gaussian distributions of 
nearest-neighbour spacings in normal conducting aluminum single 
electron transistors.  A significant part of the peak spacings is 
reduced to lower values.  We interpret this effect in terms of 
reproducible background charge rearrangements, which take place in 
close vicinity to the SET island, and are predominantly induced by 
gate voltage changes.

\section*{Acknowledgements}
Helpful discussions with A. B. Zorin, J. E. Mooij and A. Cohen are 
gratefully acknowledged. This work is supported by the Swiss Federal 
Office for Education and Science and by ETH Z\"urich.


\vskip-12pt
\begin{thebibliography}{99}
%
\bibitem{SETrev.r}
\Name{Averin D. V and Likharev K. K.} in 
\Book{Mesoscopic Phenomena in Solids} edited by 
\Name{Altshuler B. A., Lee P. A. and Webb R. A.}
(Elsevier, Amsterdam) \Year {1991}, p\Page{173-271}; \\
%
\Book{Single Charge Tunneling} edited by 
\Name{Grabert H. and Devoret M. H.} NATO ASI Series B, 
\Vol{294} (Plenum, New York) \Year {1992}.
%
\bibitem{qdot.r}
\Name{Kouwenhoven L. P. \textit{et al.}} in
\Book{Mesoscopic Electron Transport} edited by 
\Name{Sohn L. L., Kouwenhoven L. P. and Sch\"on G.} NATO ASI  
Series E, \Vol{345} (Kluwer, Dordrecht) \Year {1997}, p\Page{105-214}.
%
\bibitem{RMT.r} 
\Name{Beenakker C. W. J.} 
\Review{Rev. Mod. Phys.} \Vol{69} \Year{1997} \Page{731}.
%
\bibitem{expNNS.r}
\Name{Sivan U. \textit{et al.}} 
\Review{Phys. Rev. Lett.} \Vol{77} \Year{1996} \Page{1123};
\Name{Simmel F. \textit{et al.}} 
\Review{Europhys. Lett.} \Vol{38} \Year{1997} \Page{123};
\Name{Patel S. R. \textit{et al.}}
\Review{Phys. Rev. Lett.} \Vol{80} \Year{1998} \Page{4522}.
%
\bibitem{simmel2.r} 
\Name{Simmel F. \textit{et al.}} 
\textit{Phys. Rev.} B \textbf{59} \Year{1999} \Page{10441}.
%
\bibitem{maurer.r}
\Name{Maurer S. M. \textit{et al.}}
\Review{Phys. Rev. Lett.} \Vol{83} \Year{1999} \Page{1403}.
%
\bibitem{Fulton.r}
\Name{Fulton T.A. and Dolan G. J.} 
\Review{Phys. Rev. Lett.} \Vol{59} \Year{1987} \Page{109}.
%
\bibitem{ebox.r} 
\Name{Lafarge P. \textit{et al.}} 
\textit{Z. Phys.} B \textbf{85} \Year{1991} \Page{327}.
%
\bibitem{lt22miha.r}
\Name{Furlan M. \textit{et al.}}
to be published in \Review{Physica B} \Year{April 2000}.
%
\bibitem{SETnoise.r}
\Name{Zimmerli G. \textit{et al.}}
\Review{Appl. Phys. Lett.} \Vol{61} \Year{1992} \Page{237};
\Name{Zorin A. B. \textit{et al.}} 
\textit{Phys. Rev.} B \textbf{53} \Year{1996} \Page{13682};
\Name{Tavkhelidze A. N. and  Mygind J.}
\Review{J. Appl. Phys.} \Vol{83} \Year{1998} \Page{310};
\Name{Krupenin V. A. \textit{et al.}} 
\Review{J. Appl. Phys.} \Vol{84} \Year{1998} \Page{3212}.
%
\bibitem{kulik.r}
\Name{Kulik I. O. and Shekhter R. I.} 
\Review{Zh. Eksp. Teor. Fiz.} \Vol{68} \Year{1975} \Page{623} 
[\Review{Sov. Phys. JETP} \Vol{41} \Year{1975} \Page{308}].
%
\bibitem{chalmers99.r}
\Name{Furlan M. \textit{et al.}} 
submitted to \textit{J. Low Temp. Phys.} 
%
\bibitem{RTNrev.r}
For reviews, see \textit{e.g}
\Name{Kirton M. J. and Uren M. J.} 
\Review{Adv. Phys.} \Vol{38} \Year{1989} \Page{367}; \\
\Name{Kogan Sh.}
\Book{Electronic Noise and Fluctuations in Solids} 
(Cambridge University Press) \Year {1996}.
%
\bibitem{imagecharge.r}
\Name{Jackson J. D.}
\Book{Classical Electrodynamics} 
(2nd edition, Wiley, New York) \Year {1975}, chapter 2.
%
\bibitem{dotnoise.r}
\Name{Ralls K. S. \textit{et al.}} 
\Review{Phys. Rev. Lett.} \Vol{52} \Year{1984} \Page{228};
\Name{Cobden D. H. \textit{et al.}}
\textit{Phys. Rev.} B \textbf{44} \Year{1991} \Page{1938};
\Name{Sakamoto T. \textit{et al.}} 
\Review{Jpn. J. Appl. Phys.} \Vol{34} \Year{1995} \Page{4302}; 
\Review{Superlatt. Microstructures} \Vol{23} \Year{1998} \Page{413};
\Name{Peters M. G. \textit{et al.}}
\Review{J. Appl. Phys.} \Vol{86} \Year{1999} \Page{1523}. 
%
\bibitem{2DEGchaos.r}
\Name{Yoo M. J. \textit{et al.}} 
\Review{Science} \Vol{276} \Year{25 April 1997} \Page{579}; 
\Name{Tessmer S. H. \textit{et al.}} 
\Review{Nature} \Vol{392} \Year{5 March 1998} \Page{51}; 
\Name{Yacoby A. \textit{et al.}} 
\Review{Solid State Commun.} \Vol{111} \Year{1999} \Page{1}. 
%
\bibitem{cobden2.r}
\Name{Cobden D. H. \textit{et al.}} 
\Review{Phys. Rev. Lett.} \Vol{82} \Year{1999} \Page{4695}.
%
\end{thebibliography}
\end{document}